\RequirePackage{fix-cm}
\documentclass[smallextended]{svjour3}       
\smartqed  
\usepackage{graphicx}
\usepackage{booktabs}
\usepackage{longtable}
\usepackage{lscape}
\usepackage{graphicx, graphics}
\usepackage{float}
\usepackage{multirow}
\usepackage{bigstrut}
\usepackage{url}

\usepackage{caption}

\begin{document}
\title{Different patterns of social closeness observed in mobile phone communication}


\author{Mikaela Irene D. Fudolig \and
        Daniel Monsivais \and
        Kunal Bhattacharya \and
        Hang-Hyun Jo \and
        Kimmo Kaski
}


\institute{
Mikaela Irene D. Fudolig \at
Asia Pacific Center for Theoretical Physics, Pohang, Republic of Korea \\
\email{mikaela.fudolig@apctp.org}
          \and
          Daniel Monsivais \at
              Department of Computer Science, Aalto University, Espoo, Finland \\
              \email{monsivais.daniel@gmail.com}
          \and
          Kunal Bhattacharya \at
              Department of Computer Science, Aalto University, Espoo, Finland \\
              \email{kunal.bhattacharya@aalto.fi}
          \and
        Hang-Hyun Jo \at
          Asia Pacific Center for Theoretical Physics, Pohang, Republic of Korea \\
          Department of Physics, Pohang University of Science and Technology, Pohang, Republic of Korea \\
        Department of Computer Science, Aalto University, Espoo, Finland \\
          \email{hang-hyun.jo@apctp.org}
        \and
        Kimmo Kaski \at
            Department of Computer Science, Aalto University, Espoo, Finland \\
            The Alan Turing Institute, London, UK \\
            \email{kimmo.kaski@aalto.fi}
}

\date{Received: date / Accepted: date}

\maketitle

\begin{abstract}
We analyze a large-scale mobile phone call dataset containing information on the age, gender, and billing locality of users to get insight into social closeness in pairs of individuals of similar age. We show that in addition to using the demographic information, the ranking of contacts by their call frequency in egocentric networks is crucial to characterize the different communication patterns. We find that mutually top-ranked opposite-gender pairs show the highest levels of call frequency and daily regularity, which is consistent with the behavior of real-life romantic partners. At somewhat lower level of call frequency and daily regularity come the mutually top-ranked same-gender pairs, while the lowest call frequency and daily regularity are observed for mutually non-top-ranked pairs. We have also observed that older pairs tend to call less frequently and less regularly than younger pairs, while the average call durations exhibit a more complex dependence on age. We expect that a more detailed analysis can help us better characterize the nature of relationships between pairs of individuals and distinguish between various types of relations, such as siblings, friends, and romantic partners.

\end{abstract}

\section{Introduction}\label{sec:intro}

Traditionally, the studies of human relationships and human social networks have been conducted using questionnaire-based surveys~\cite{Zachary1977AnGroups,Wasserman1994SocialAnalysis}. As these surveys focus on detailed information about social ties between human individuals, they tend to be limited by the number of subjects, by the relative uniformity of subjects often recruited from the same social surrounding, and by the memory of the subjects filling the questionnaires. Recent digital technologies such as mobile phones and Radio Frequency Identification (RFID) have enabled researchers to supplement the survey data with much more detailed relational data between subjects~\cite{Eagle2006RealitySystems,Jo2012SpatiotemporalUsages,Cattuto2010DynamicsNetworks}. Although the data in these studies are still limited in sample size and in the diversity of the subjects, they have paved the way for more accurate and quantitative description of social behavior of human individuals embedded in a social environment or network. An additional benefit of this type of studies is that they allow the cross-validation of the data gathered from different sources~\cite{Stopczynski2014MeasuringResolution,Wang2014StudentLife}. 

Large-scale mobile phone datasets have also become available due to rapid advance of digital mobile phone technology in the hands of people generating vast amount of data as traces of their behavior. This has facilitated a complementary approach to investigate human sociality even at the population level. The \emph{call detail records} (CDRs) of mobile phone usage enable us to map the patterns of sociality at diverse scales, i.e., from studying the structure and dynamics of large-scale social networks~\cite{Onnela2007StructureNetworks,Karsai2011SmallSpreading}, to the level of communities and groups~\cite{Blondel2008FastNetworks}, to immediate social neighborhood of individuals in terms of egocentric networks~\cite{Palchykov2012SexRelationships,Jo2014SpatialLifespan}. In these studies, the strength of ties between individuals has often been quantified in terms of the frequency of contact between them. More recently, such information on the tie strength has been combined with the metadata, such as the age, gender, and billing post code of users, which in turn has enabled us to gain deeper insight into the nature of human sociality~\cite{Palchykov2012SexRelationships,Jo2014SpatialLifespan,Bhattacharya2016SexHumans,Dong2014InferringNetworks}.

However, as the CDRs are anonymized, they do not carry the true nature of relationships between individuals. An approach to circumvent this issue is to utilize the demographic and/or location information of the users and to make plausible assumptions about the nature of relationships between the users~\cite{David-Barrett2016CommunicationCourse,David-Barrett2017PeerFormation}. For example, for a given user (an ego), the contacts of the ego (alters) are ranked by the call frequency between the ego and each alter; a few of the top-ranked alters are selected for the study. Then the tie strengths of close relationships are correlated with the age, gender, and location information of the users. The findings in these studies turn out to be indicative and consistent with the well-understood life-course patterns of human sociality~\cite{Hawkes2004TheEffect,Kalmijn2012LongitudinalSupport}. However, since this approach only uses demographic information of the users to infer the relationship between them, it cannot distinguish between pairs with similar demographics but different relationships, e.g., between opposite-gender friends and opposite-gender romantic partners. To address this limitation, we study whether different communication patterns can be found even among pairs with similar demographics. If pairs with similar demographic information exhibit different communication patterns, it may indicate that these pairs correspond to different types of relationships.

In the present study, we extend the above-described approach to analyze the CDRs focusing on getting insight into the nature of relationships between pairs of individuals of similar age. We combine the metadata of users, including age, gender, and billing post code, with the information about the ranks in each other's egocentric networks to find groups of pairs showing different calling patterns in terms of the average daily call frequency, daily regularity, and average call duration. These groups can be interpreted as corresponding to either ``close'' or ``casual'' pairs, with close pairs being characterized by higher call frequency and regularity as well as longer call durations than those of casual pairs. We find that the rank information, which has not been used in any previous work, is crucial to separate these groups. We note that although some socially close pairs do not necessarily call frequently, it has been shown that calling patterns can be reliably used to predict user-validated social closeness~\cite{Phithakkitnukoon2010}.

Our paper is organized as follows: in Sect.~\ref{sec:methods} we present the description of the data used in this study, followed by the methods of data preprocessing and statistical tests. Then in Sect.~\ref{sec:results} we focus on distinguishing the pairs of individuals that show different calling patterns in terms of average daily call frequency, daily regularity, and average call duration. In Sect.~\ref{sec:discussion} we discuss how these results can be used to interpret pairs as being either close or casual. We also relate our findings to those from small-scale sociological studies. Finally, we draw conclusions in Sect.~\ref{sec:conclusion}.

\section{Methods}\label{sec:methods}
\subsection{Data description}
\label{subsec:data_desc}

We analyze the mobile phone call dataset of a European service provider for the first 7 months of year 2007 (212 days).
During this period, which is before the rise of smartphones and social network services, a significant part of the mobile communication was done through voice calls and Short Message Services (SMSs). The service provider had subscribers numbering around 20\% of the population of the country~\cite{Onnela2007StructureNetworks}. 

The dataset contains the date and time for all the outgoing and incoming calls between subscribers or users. The duration is included for the calls between the users and for the outgoing calls from the users to those who subscribed to other providers, which we call non-company users. The duration is zero for incoming calls from non-company users to users, but the date and time of such calls are included. We discard the users whose contracts are known to begin or end within the period of interest, i.e., the first seven months in 2007.

For each service contract, some metadata of the users, such as age and gender, are included in the dataset; for most users, the billing post code is also included. There are a number of users, however, with unknown or ambiguous age and gender: these include the non-company users who have no user metadata, as well as users associated with service contracts that have multiple identifiers. Since it is impossible to determine whether the multiple identifiers correspond to different phones owned by the same person or different individuals, only the identifier with the earliest subscription date retained the age and gender information, while other identifiers are considered to have unknown age and gender.

\subsection{Data preprocessing}
\label{subsec:preprocessing}

For each user with known metadata, which we call an ``ego'', we enlist all the other users the ego communicated with, which we call ``alters''. The alters, which may include users with unknown age and gender, are ranked in descending order according to the total number of incoming and outgoing calls made between the ego and each alter. By keeping the top five alters for each ego, which likely correspond to the ego's emotionally closest alters~\cite{Dunbar1995,Zhou2005}, we make the list of ego--alter pairs.

We are interested only in ego--alter pairs who have significant relationships, such as family, friends, and romantic partners. To filter out pairs which do not meet this criterion, we impose regularity by excluding purely transactional calls, which are characterized by lower call frequency and less regularity~\cite{Granovetter1973TheTies}. Specifically, we exclude pairs who have had calls in less than five out of the seven months. For example, if a pair has one thousand calls but only for the first month, we exclude that pair from the analysis. Further, we also exclude the ego--alter pairs in which the metadata of the alter does not contain the age and gender, including the non-company users. After these filtering steps, we are left with the users with known metadata who make calls regularly to each other. Note that, since the filtering is done after ranking, the ranking preserves the true importance of the alter to the ego, as far as call frequency is concerned.

It is possible that two users appear in each other's list of the top five alters. In such a case, the total number of incoming and outgoing calls and the total call duration are the same for both users, but this pair appears twice in the list of ego--alter pairs; we keep only one of these two. 

\begin{table}[t!]
\centering
\caption{Numbers of pairs in nine different demographic groups according to the genders and the younger user's age in each pair, with percentages in each group when decomposed by the ranks of users in each other's egocentric networks, i.e., mutual top-rank (1-1), mutual non-top-rank (n-n), and non-mutual top-rank (1-n). Note that due to rounding errors, the percentages may not sum to exactly 100\%.}
\label{table:user-counts}
\resizebox{\textwidth}{!}{
\begin{tabular}{@{}ccccccc@{}}
\toprule

 & \multicolumn{2}{c}{\textbf{Opposite gender ($-$)}} & \multicolumn{2}{c}{\textbf{Same-gender female (+f)}} & \multicolumn{2}{c}{\textbf{Same-gender male (+m)}} \\ 

\midrule
\textbf{Young adulthood} & \multicolumn{2}{c}{95,128} & \multicolumn{2}{c}{47,930} & \multicolumn{2}{c}{70,945} \\ \cmidrule(l){2-7}
\textbf{(Y)} & 1-1 & 38.0\% & 1-1 & 6.9\% & 1-1 & 11.0\% \\
 & n-n & 45.6\% & n-n & 75.5\% & n-n & 69.9\% \\
 & 1-n & 16.5\% & 1-n & 17.7\% & 1-n & 19.2\% \\

\midrule
\textbf{Middle adulthood} & \multicolumn{2}{c}{232,133} & \multicolumn{2}{c}{98,225} & \multicolumn{2}{c}{120,256} \\ \cmidrule(l){2-7}
\textbf{(M)} & 1-1 & 40.3\% & 1-1 & 9.9\% & 1-1 & 14.0\%\\
 & n-n & 38.5\% & n-n & 68.8\% & n-n & 65.0\%\\
 & 1-n & 21.2\% & 1-n & 21.3\% & 1-n & 21.0\%\\

\midrule
\textbf{Late adulthood} & \multicolumn{2}{c}{57,085} & \multicolumn{2}{c}{20,165} & \multicolumn{2}{c}{26,544} \\ \cmidrule(l){2-7}
\textbf{(L)} & 1-1 & 30.1\% & 1-1 & 14.7\% & 1-1 & 19.9\%\\
 & n-n & 40.0\% & n-n & 59.3\% & n-n & 50.6\%\\
 & 1-n & 29.9\% & 1-n & 26.0\% & 1-n & 29.5\%\\
\bottomrule
\end{tabular}
}
\end{table}

After the above-described ranking, filtering, and removing of duplicates, we are left with 322,823 users in 1,236,364 pairs. Of these, we consider the pairs more likely to be in a peer relationship rather than in a parent--child relationship. These two relationships can be distinguished using the age difference of the users; we set the cutoff to be 20 years, based on European census data~\cite{EurostatEurostatDatabase}. Then we study pairs whose age difference is less than 20 years, which are then categorized into nine demographic groups. We first consider three combinations of genders of each pair: (1) opposite gender, denoted by ``$-$'', (2) same-gender female or ``+f'', and (3) same-gender male or ``+m''. For each gender combination group, we consider three age groups according to the age of the younger user in a pair, being either 18--28 years old (young adulthood or ``Y''), 29--45 years old (middle adulthood or ``M''), or 46--55 years old (late adulthood or ``L''), following the scheme of life stages used in Ref.~\cite{David-Barrett2016CommunicationCourse}.\footnote{Other age categorization schemes were tried, such as considering only pairs where both users belong to the same age group or categorizing pairs based on the average age of the two users. The results are found to be robust even with these changes in the categorization scheme.} Consequently, we focus on 768,411 pairs in nine demographically separable groups, denoted by $-$Y, $-$M, $-$L, +fY, +fM, +fL, +mY, +mM, and +mL, respectively. Each of these groups has at least 20,000 ego--alter pairs, as summarized in Table~\ref{table:user-counts}. Also, although we consider the maximum age difference to be 20 years, we find that most of these pairs show an age difference of only 0--5 years, as shown in Table~\ref{table:age-dist}.

\begin{table}[t!]
\centering
\caption{Distributions of the age differences of the pairs. For all demographic groups, pairs with an age difference of 0--5 years comprise more than half of each group, followed by those with age differences in the range of 6--10 years. Although the peers are defined to have an age difference less than 20 years, most of the pairs in each group show age differences less than 10 years. Due to rounding errors, the percentages may not sum to exactly 100\%.}
\label{table:age-dist}
\resizebox{\textwidth}{!}{
\begin{tabular}{@{}lcccccccccc@{}}

\toprule
 & Age difference & \multicolumn{3}{c}{\textbf{Opposite-gender ($-$)}} & \multicolumn{3}{c}{\textbf{Same-gender female (+f)}} & \multicolumn{3}{c}{\textbf{Same-gender male (+m)}} \\ \cmidrule(l){3-11}
 & (years)  & 1-1 & n-n & 1-n & 1-1 & n-n & 1-n  & 1-1 & n-n & 1-n  \\ \midrule
 & 0--5 & 73\% & 56\% & 62\% & 68\% & 64\% & 65\% & 71\% & 68\% & 71\% \\
\textbf{Y} & 6--10 & 20\% & 26\% & 23\% & 17\% & 23\% & 20\% & 20\% & 21\% & 19\% \\
 & 11--19 & 7\% & 18\% & 15\% & 15\% & 13\% & 15\% & 9\% & 11\% & 10\% \\ \midrule
 & 0--5 & 78\% & 58\% & 68\% & 69\% & 64\% & 65\% & 75\% & 67\% & 69\% \\
\textbf{M} & 6--10 & 16\% & 24\% & 20\% & 16\% & 21\% & 17\% & 17\% & 21\% & 19\% \\
 & 11--19 & 6\% & 18\% & 13\% & 15\% & 15\% & 17\% & 8\% & 13\% & 12\% \\ \midrule
 & 0--5 & 76\% & 62\% & 72\% & 75\% & 61\% & 71\% & 83\% & 65\% & 76\% \\
\textbf{L} & 6--10 & 17\% & 23\% & 19\% & 15\% & 23\% & 17\% & 12\% & 21\% & 15\% \\
 & 11--19 & 6\% & 15\% & 9\% & 10\% & 16\% & 12\% & 6\% & 14\% & 8\% \\ \bottomrule 
\end{tabular}
}
\end{table}

Each demographic group can be further divided into three subgroups according to the ranks of users in each other's egocentric networks, as follows: (1) Both users in a pair are the top-rank alters of each other, which can be called \emph{mutual top-rank} and denoted by ``1-1'', (2) both users are not the top-rank alters of each other, i.e., \emph{mutual non-top-rank} or ``\emph{n}-\emph{n}'', and (3) one of the users is the top-rank alter of the other, but it is not mutual, i.e., \emph{non-mutual top-rank} or ``1-\emph{n}''. Table~\ref{table:user-counts} shows that for all age groups, mutual top-rank pairs comprise a large portion in the opposite-gender groups, while they are a small minority in the same-gender groups.

In addition to all of the above, we can extract the location information of users with the help of the billing post code, which we assume to correspond to the user's home address. We will focus on whether the users of each pair have the same post code or different ones.

\subsection{Statistical test}

All the statistical tests are done on the log-transformed variable whenever necessary. To test for statistical significance, we use one-way ANOVA and Tukey's HSD post hoc test when the variances are found equal by Levene's test~\cite{2013NIST/SEMATECHMethods}. If heteroscedasticity is obtained, Welch's ANOVA~\cite{Welch1947TheInvolved} and the Games--Howell post hoc test~\cite{Games1976PairwiseStudy} are used instead. The tests are implemented using Python's \texttt{scipy} and \texttt{statsmodels} as well as using R's \texttt{userfriendlyscience} packages.

Due to the large sample sizes in this study, the power of statistical tests is high~\cite{Columb2016StatisticalEstimations}, and true differences, no matter how small they are, are more likely to be found as significant. For brevity, we only mention the relevant results of the statistical tests where the null hypothesis cannot be rejected. Otherwise, the statistical tests either show a significant difference or are overruled by practical significance. We also report Cohen's \emph{d}, an effect size which quantifies the difference in means between two groups~\cite{Sawilowsky2009}, whenever useful (see the details in Supplementary Material).

\section{Results}\label{sec:results}
To quantify the calling patterns of ego--alter pairs, we introduce three quantities, i.e., the average daily call frequency, fraction of days active, and average call duration. The distributions of these quantities are then systematically compared across different demographic groups.
\subsection{Average daily call frequency}

\begin{figure}[t!]
 \centering
\includegraphics[width=\linewidth]{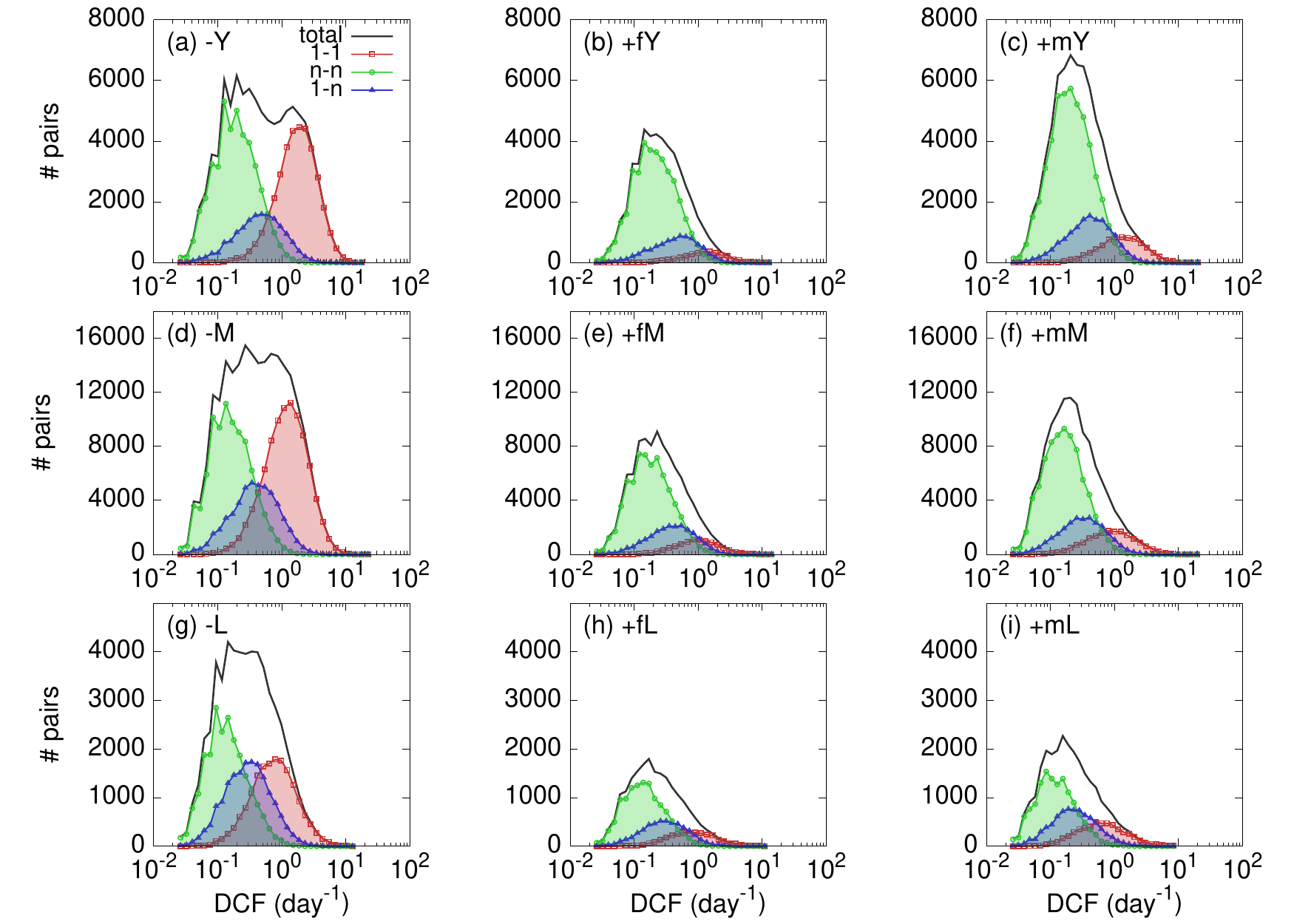}
 \caption{\textbf{Distributions of the average daily call frequency (DCF).}
For each demographic group, the total distribution (black solid line) is decomposed into three subgroups depending on the ranks of users in each other's egocentric networks: mutual top-rank (1-1; red line with squares), mutual non-top-rank (n-n; green line with circles), and non-mutual top-rank (1-n; blue line with triangles). The unit of DCF is day$^{-1}$.}
\label{fig:dcf}
\end{figure}

\begin{figure}[t!]
\centering
\includegraphics[width=0.8\linewidth]{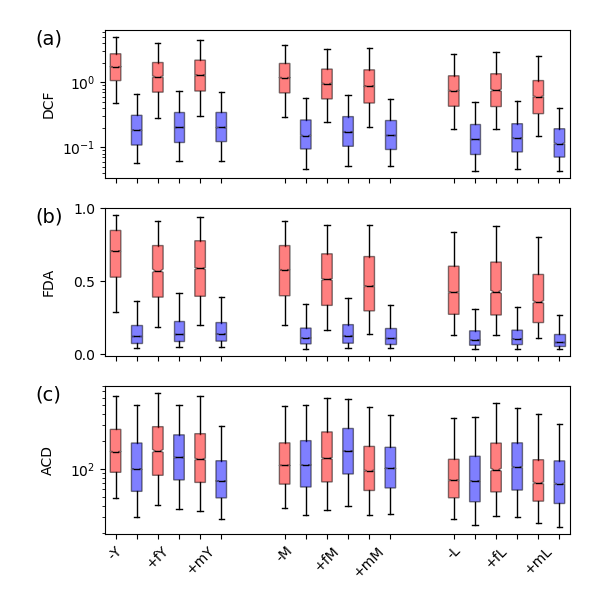}
\caption{\textbf{Summary of distributions of the average daily call frequency (DCF), the fraction of days active (FDA), and the average call duration (ACD).}
The pink bars correspond to the mutual top-rank (1-1) pairs, while the blue bars are for the mutual non-top-rank (n-n) pairs. The black lines in the middle indicate the medians, the bars include data points from the 25th to the 75th percentile, and the whiskers show the 5th and 95th percentiles. The notches show confidence intervals generated by bootstrapping with 10000 resamples. The units of DCF and ACD are day$^{-1}$ and s, respectively.}
\label{fig:boxplot}
\end{figure}

We first obtain the number of calls made by each pair, i.e., the call frequency. Dividing this call frequency by the number of days in the observation period, i.e., 212 days, we get the \textit{average daily call frequency} (DCF) per pair to obtain its distributions. In Fig.~\ref{fig:dcf}, we find that the distribution of DCFs for each of nine demographic groups can overall be described by unimodal distributions on a log-scale, except for the opposite-gender young adulthood ($-$Y) case in Fig.~\ref{fig:dcf}a, showing a clear bimodality. This bimodality is resolved by separating the pairs in the $-$Y  group according to the ranks of the users in each other's egocentric networks. We observe that the mutual top-rank (1-1) pairs and mutual non-top-rank (n-n) pairs successfully account for the right and left peaks of the bimodal distribution, such that the median values for 1-1 and n-n pairs are around $1.75$ and $0.18$ calls per day, respectively. There is also a non-mutual top-rank (1-n) minority whose distribution shows a peak between those of mutual top-rank and mutual non-top-rank pairs, which will be discussed in Subsect.~\ref{subsec:other}. Although the bimodality found in the $-$Y case is not evident in the rest of the groups, we find that the mutual top-rank pairs show, in general, largely different calling patterns from the mutual non-top-rank pairs in all the other demographic groups, as depicted in Fig.~\ref{fig:boxplot}a. Note that although mutual top-rank pairs are expected to have more calls than mutual non-top-rank pairs, the successful decomposition of the bimodality using the rank information is not straightforward. 

We summarize other relevant findings from the results in Figs.~\ref{fig:dcf}~and~\ref{fig:boxplot}a. For all the gender combinations of pairs, younger pairs tend to call considerably (slightly) more often than older pairs in the mutual top-rank (mutual non-top-rank) case, with the effect sizes most pronounced among opposite-gender peers. For the mutual top-rank case, opposite-gender pairs call more frequently than their same-gender counterparts for both Y and M groups, while for the oldest (L) groups, there is no significant difference at $\alpha = 0.05$ between opposite-gender and same-gender female pairs ($p=0.18$), but both opposite-gender and same-gender female pairs call more often than the same-gender male pairs. On the other hand, for the mutual non-top-rank case, we find no clear gender dependence of the DCF for each age group.

\subsection{Daily regularity}

To quantify the temporal regularity of the calling patterns on a daily basis, we define the \emph{fraction of days active} (FDA) as the fraction of days in the observation period in which at least one call was made between the users of each pair. By the FDA, one can distinguish, e.g., the case of 10 days with 10 calls per day from the case of 100 days with one call per day, which cannot be distinguished by the average daily call frequency (DCF).

\begin{figure}[t!]
\centering
\includegraphics[width=\linewidth]{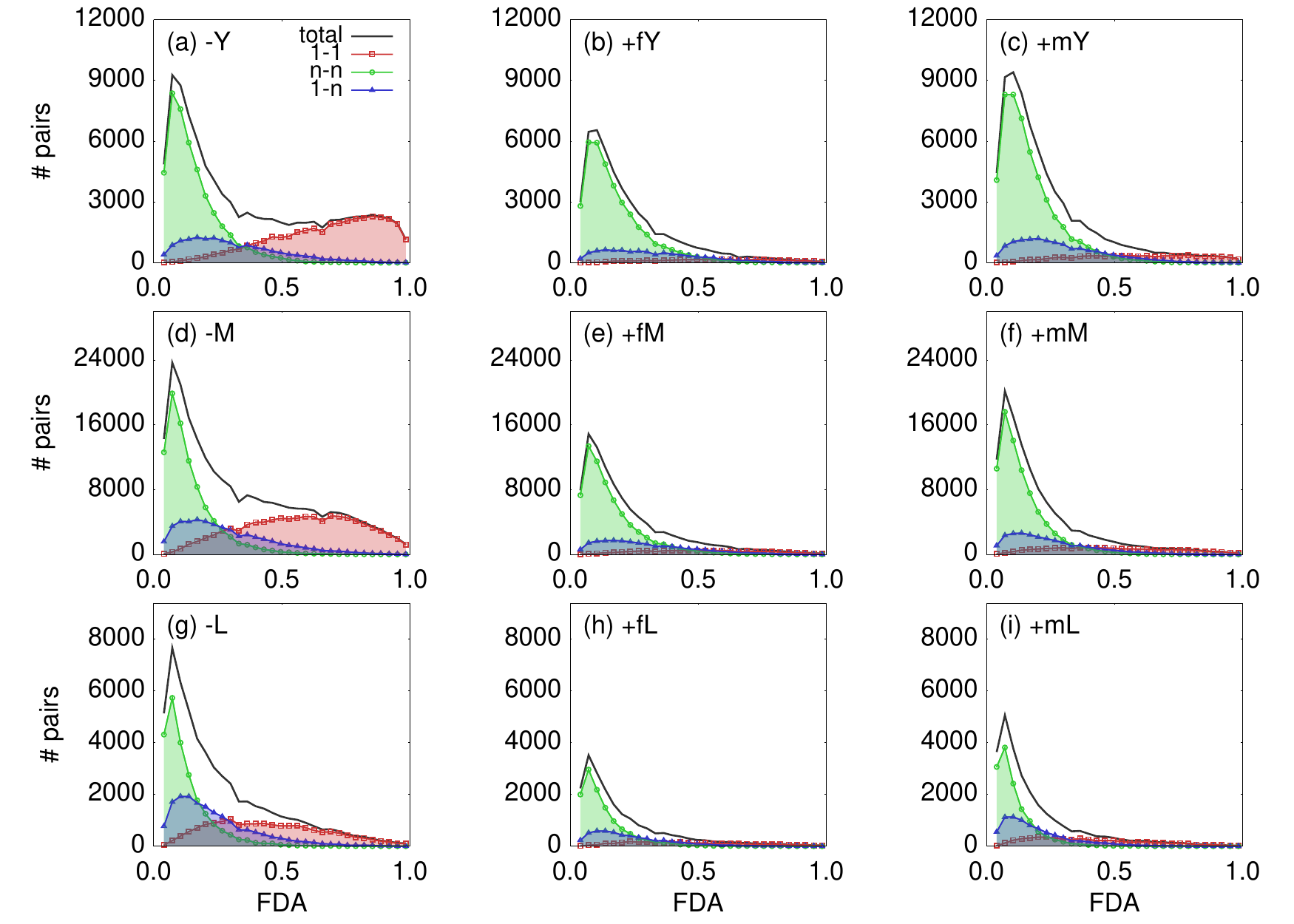}
 \caption{\textbf{Distributions of the fraction of days active (FDA).}
 All notations are the same as those in Fig.~\ref{fig:dcf}.
 }
\label{fig:fda}
\end{figure}

We find that the FDA is highly correlated with the DCF ($r=0.681$). It should be noted that the number of days active cannot be greater than the call frequency for each pair, which possibly enables the strongly positive correlation between FDA and DCF. However, how the calls are distributed over the observation period is yet an interesting question, in particular, for pairs with high DCF: the pairs with a high DCF tend to have a high FDA, implying that the calls are made rather regularly instead of being lumped into a few days. The distributions of FDAs are presented in Fig.~\ref{fig:fda} and summarized in Fig.~\ref{fig:boxplot}b. Overall, we find similar behavior to that observed in the case of DCF, except that the shapes of the distributions are highly skewed either to the left or to the right, probably due to the intrinsic range of the quantity, i.e., FDA~$\in[0,1]$. The most pronounced difference between the mutual top-rank and the mutual non-top-rank pairs is observed again in the $-$Y group as their median values are $0.71$ and $0.13$, respectively.

\subsection{Average call duration}

\begin{figure}[t!]
\centering
\includegraphics[width=\linewidth]{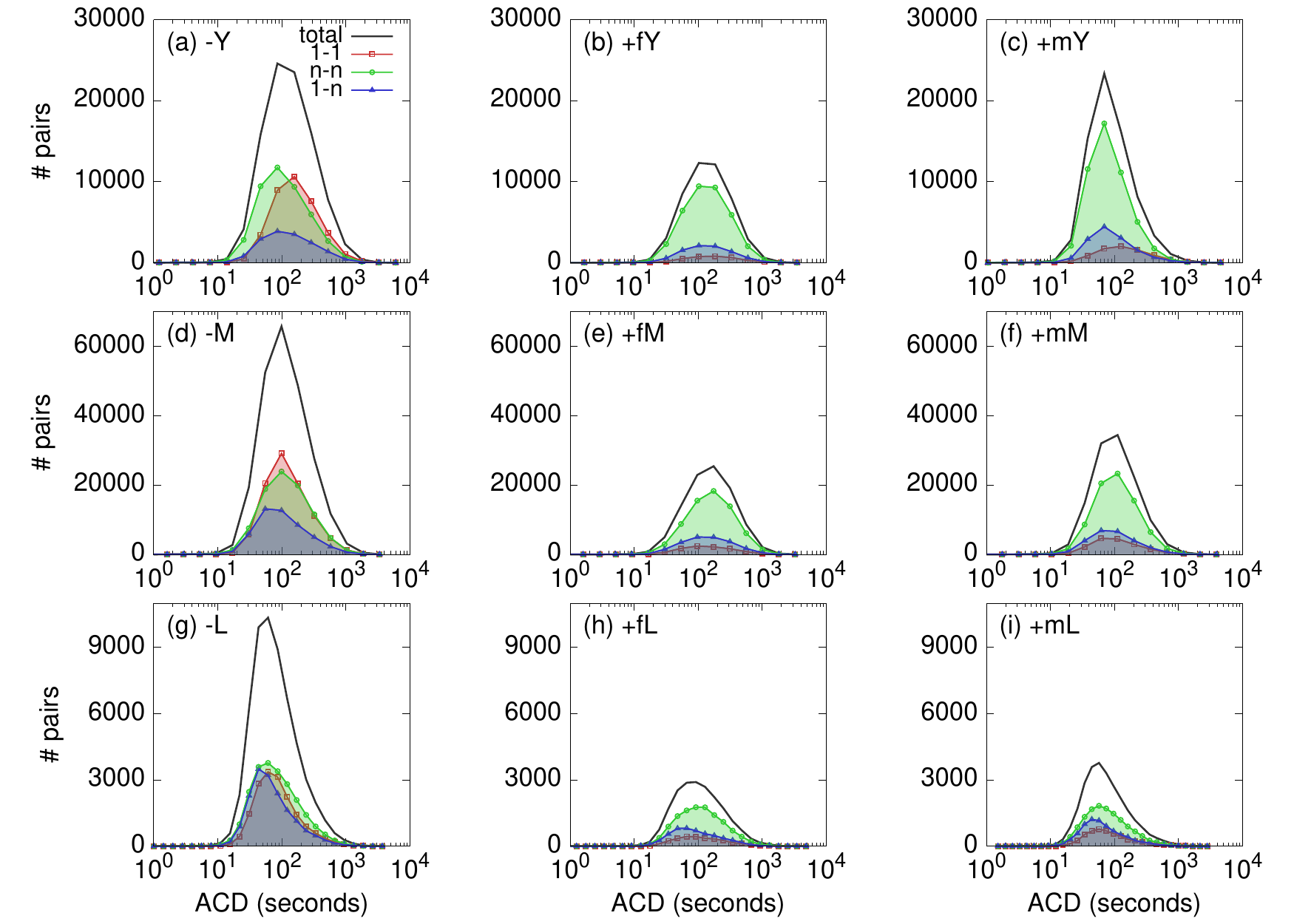}
 \caption{\textbf{Distributions of the average call duration (ACD).}
 All notations are the same as those in Fig.~\ref{fig:dcf}, except that the unit of ACD is seconds.}
 \label{fig:acd}
\end{figure}

To study the calling patterns in more detail, we calculate the average duration per call or the \textit{average call duration} (ACD) for each pair by dividing the total call duration (in seconds) by the number of calls. The ACD turns out to be positively but weakly correlated with the DCF ($r=0.087$) as well as with the FDA ($r=0.075$). As shown in Figs.~\ref{fig:boxplot}c and~\ref{fig:acd}, unlike the DCF and FDA, there seems to be no clear demographic dependence of the ACD across the different age and gender groups. However, with the medians of the distributions we observe that for all the gender groups, younger pairs tend to have longer calls than older pairs only in the mutual top-rank case. Interestingly, the same-gender female pairs make longer calls than their opposite-gender and same-gender male counterparts for all the age groups, regardless of the ranks, except for one case; there is no significant difference between the mutual top-rank pairs in $-$Y and +fY groups ($p=0.79$). We also find that in the young adulthood (Y) case, the mutual top-rank pairs have longer calls than the mutual non-top-rank pairs for all gender combinations, while the opposite tendency is significantly observed for the +fM group.

\subsection{Location analysis}

\begin{figure}[t!]
 \centering
\includegraphics[width=0.9\linewidth]{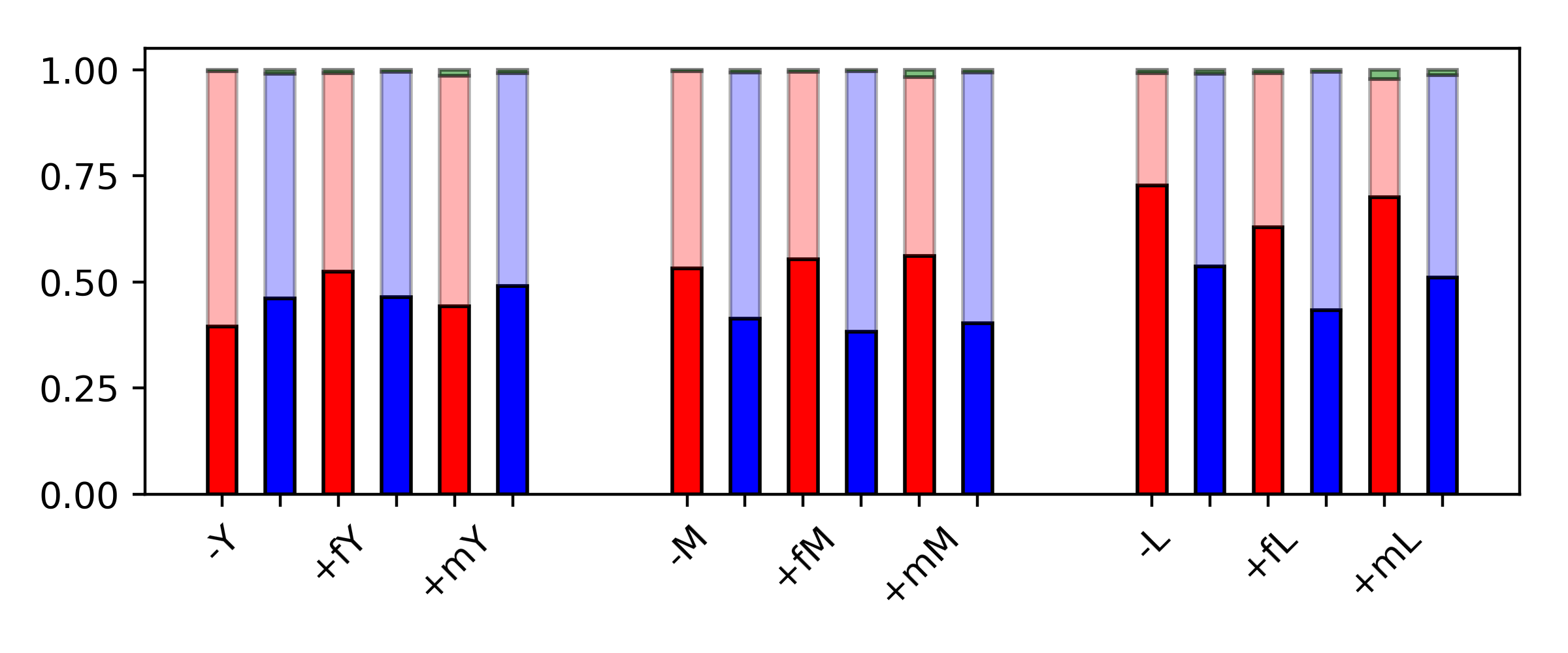}
 \caption{\textbf{Decomposition of groups by the billing post code.}
The red bars correspond to mutual top-rank (1-1) pairs, while the blue bars correspond to mutual non-top-rank (n-n) pairs for each group. The darker bars denote the fraction of pairs with the same post codes, while the lighter bars denote the fraction of those with different post codes. The fractions of pairs in which at least one post code is not available are denoted by the green bars.
 }
 \label{fig:post-code-bar}
\end{figure}

Next we analyze the location information of the pairs in close relationships by assuming that the billing post codes correspond to the home address of the users. It is known that the frequency of the face-to-face interaction, constrained by the location, is positively correlated with the frequency of contact by telephone and other media~\cite{Roberts2009ExploringCharacteristics,Roberts2011CommunicationCloseness,Saramaki2014PersistenceCommunication}. This enables us to study how the locations of users in close relationships are related to each other. In Fig.~\ref{fig:post-code-bar}, we find that although the fraction of pairs living in the same post code increases with age of the group for all mutual top-rank pairs, the difference is most pronounced among opposite-gender pairs. 

\section{Discussion}\label{sec:discussion}

Based on the above empirical observations, we hypothesize that across all demographic groups, mutual top-rank pairs and mutual non-top-rank pairs have essentially different calling patterns, implying different types of relationships that underlie the communication patterns. Mutual top-rank (1-1) pairs are characterized by a high number of calls and high daily regularity, and hence these relationships can be interpreted as close. On the other hand, mutual non-top-rank (n-n) pairs have fewer calls and very low regularity, leading us to interpret these relationships as casual. As the calling patterns of mutual top-rank pairs are also differentiated by their genders, in the following we discuss three types of relationships: close relationships of opposite-gender pairs, close relationships of same-gender pairs, and casual relationships.

\subsection{Close relationships of opposite-gender pairs}

The opposite-gender, mutual top-rank pairs show the highest level of call frequency and regularity compared to all other gender and rank cases. This indicates a high level of closeness and suggests that these relationships are possibly romantic in nature. Although this interpretation cannot be confirmed by ground truth, our findings are consistent with those of small-scale sociological studies involving college students (corresponding to the age group of Y) in confirmed romantic relationships~\cite{Jin2010MobileStyles,Morey2013YoungStyle}. These studies find that pairs with greater frequency or duration of phone calls have less relational uncertainty and higher intimacy. Moreover, those in romantic relationships are, on average, found to call each other more regularly~\cite{Morey2013YoungStyle}. In addition, we find that the mutual top-rank pairs form a significant chunk in the opposite-gender groups, but only a small minority in the same-gender groups. Since this reflects the profile of marriages and registered partnerships in a number of countries in Europe, where same-sex unions account for less than 5\% of all registered unions~\cite{StatistischesBundesamt,InstitutoNacionaldeEstadistica,OfficeforNationalStatistics}, romantic partnerships seem to be the most plausible characterization of these opposite-gender pairs.

We also observe that for this kind of relationship, younger pairs tend to have more frequent, more regular, and longer calls than older pairs, as depicted in Fig.~\ref{fig:boxplot}. To study whether this tendency is due to the lower usage of mobile phones among the older generation~\cite{Kurniawan2008OlderInvestigation} or due to the actual communication patterns of older users, more work is called for.

Recall from Fig.~\ref{fig:post-code-bar} that the fraction of pairs living in the same post code increases with age of the group for all mutual top-rank pairs, especially among opposite-gender pairs. If these opposite-gender pairs are romantic, this may be explained by the higher likelihood of cohabitation and/or marriage as people get older, which is consistent with the previous empirical findings using the same dataset~\cite{Jo2014SpatialLifespan}.

\subsection{Close relationships of same-gender pairs}

The same-gender, mutual top-rank pairs can also be considered as being close across all age groups as their calling patterns are clearly more active and regular than their mutual non-top-rank counterparts. Yet, they are less active and less regular than their opposite-gender counterparts, which implies that the same-gender, mutual top-rank pairs have a different type of relationship compared to close opposite-gender pairs. However, such differences turn out to get smaller for older age groups. In addition, in terms of the median of the distribution of average call duration, the same-gender female pairs tend to have longer calls than the opposite-gender and same-gender male pairs, which is consistent with the findings in Refs.~\cite{Bhattacharya2016SexHumans,David-Barrett2016CommunicationCourse}.

To characterize some of the same-gender close relationships as romantic, we need more supporting evidence for the communication patterns of same-gender romantic relationships. The most one can say is that these pairs may be a mixture of romantic, familial, and other relationships.

\subsection{Casual relationships}

The mutual non-top-rank pairs in all demographic groups are here considered as casual relationships, as they are characterized by the lowest level of call frequency and daily regularity compared to their mutual top-rank counterparts. As for the average call duration, in terms of the median of its distribution, the mutual non-top-rank casual pairs tend to have shorter or similar call durations than the mutual top-rank close pairs in most cases, except for the +fM case, where the average call duration of casual pairs (around 157 s) is significantly larger than that of the close pairs (around 134 s). Moreover, among the mutual non-top-rank pairs, the +fM group shows the longest average call duration, which could be due to requirements of child rearing, job demands in the mid- to high-level careers, or other life events.

Similarly to the opposite-gender and same-gender close relationships, the average daily call frequency and the fraction of days active in casual relationships are decreasing with their age. However, the average call duration is the highest for the middle adulthood (M) group irrespective of gender. 

We also note that unlike for the close pairs, the fraction of casual pairs with the same billing post code does not show a clear dependence on age group as is evident in Fig.~\ref{fig:post-code-bar}.

\subsection{Other relevant issues}\label{subsec:other}

So far we have focused on the ego--alter pairs as if they are separated from the rest of the social network. By incorporating the network structure surrounding those pairs, one can tackle some unresolved issues. For example, friends, family, and romantic relationships may be differentiated using the information about their common contacts, while the non-mutual top-rank (1-n) pairs may be studied in the context of directed relationships~\cite{Ball2013FriendshipStatus,Almaatouq2016AreChange}. Since the peak of the DCF distribution in the 1-n case lies between the mutual top-rank (1-1) and mutual non-top-rank (n-n) peaks, we can hypothesize that they may exhibit different behaviors from both. They may also be composed of two subgroups, one resembling the mutual top-rank pairs, the other resembling the mutual non-top-rank pairs.

\section{Conclusion}\label{sec:conclusion}

We have analyzed the large-scale call detail records (CDRs) with the metadata, such as the age, gender, and billing post code of mobile phone users, by focusing on around 770,000 pairs with an age difference of less than 20 years. We show that in addition to the metadata, the ranks of users in each other's egocentric networks, determined by the call frequency between them, reveal different communication patterns even among pairs of individuals with similar demographic information. Whereas previous studies have used only demographic information to determine the most likely relationship between users~\cite{David-Barrett2016CommunicationCourse,David-Barrett2017PeerFormation}, our findings show that adding rank information may give us insights into how the nature of relationships can vary even within the same demographic group. In particular, mutual top-rank pairs have markedly different calling patterns from mutual non-top-rank pairs, not only in terms of call frequency but also in terms of daily regularity. These differences could enable us to interpret mutual top-rank pairs as being close relationships and mutual non-top-rank pairs as being casual relationships, respectively. Although there is no information on the actual nature of the relationships among the communicating individuals, our interpretation is consistent with the findings from small-scale sociological studies with verified ground truth.

We have found that mutual top-rank pairs are much more common among opposite-gender pairs. This, as well as the consistency of their calling patterns with those observed for romantic couples, makes it plausible that mutually top-ranked opposite-gender pairs reflect romantic relationships. On the other hand, although mutually top-ranked same-gender pairs also have relatively high call frequency and regularity, they have different calling patterns compared to their mutually top-ranked opposite-gender counterparts. This may be because the same-gender group is not solely composed of romantic partnerships, but may also include platonic or familial ties as well. In contrast to the mutual top-rank pairs, the mutual non-top-rank pairs exhibit the lowest levels of daily regularity and call frequency. We suppose that these pairs are very unlikely to be romantic partners, but instead they are more likely to be platonic or familial pairs.

The calling patterns between pairs of individuals have also been found to vary with the age of the users. We find that older pairs tend to call less frequently and less regularly than younger pairs. The mutual top-rank pairs tend to also have longer average call durations than the mutual non-top-rank pairs for younger pairs of 18--28 years old. For the older pairs, the difference between the mutual top-rank and the mutual non-top-rank pairs is smaller. Interestingly, we find that in the case of the female pairs in the age range of 29--45 years, the mutual non-top-rank pairs make significantly longer calls than the mutual top-rank pairs. This age range corresponds to the period when most couples begin families; hence, such calling patterns may be due to the demands of family and work on women of that age range. Our findings can be related to the shift in social focus: while both men and women in their young adulthood are likely to maintain stronger social focus on their partners~\cite{Burton-Chellew2015RomanceCostly}, the attention of individuals in middle adulthood gets distributed to alters other than their partners due to time constraints and the increase in the number of familial ties~\cite{Bhattacharya2016SexHumans}.

We also comment on the limitations of our work. First, although non-company users were included in the ranking of the alters by call frequency, they could not be analyzed in more detail as their age and gender were unknown. Thus, our work is biased towards pairs that belong to the same service provider. Second, our dataset was extracted in the year 2007 before the rise of internet-based messaging applications; thus, we expect our dataset to cover most mobile communication at that time, which is via calls and texts. However, advances in technology may have changed the way people communicate, and it would be interesting to apply our methods to more current datasets if they are made available by the application developers.

Finally, we discuss possible future studies. While we have focused exclusively on peers, we can also investigate the parent--child relationships. In addition, as our analysis has focused on the ego--alter pairs, network analysis may help us to uncover more about the users' relationships, and even to distinguish between other types of relationships, such as platonic and familial relationships. These are all interesting issues for future work.

\begin{acknowledgements}
We acknowledge Takayuki Hiraoka for helpful discussions and we also acknowledge the computational resources provided by the Aalto Science-IT project.
    D.M. acknowledges CONACYT, Mexico for supporting grant 383907.
    D.M., K.B., and K.K. acknowledge support from EU HORIZON 2020 FET Open RIA project (IBSEN) No. 662725, and EU HORIZON 2020 INFRAIA-1-2014-2015 program project (SoBigData) No. 654024. 
    K.K. also acknowledges the Rutherford Foundation Visiting Fellowship at The Alan Turing Institute, UK. 
    H.-H.J. acknowledges financial support by Basic Science Research Program through the National Research Foundation of Korea (NRF) grant funded by the Ministry of Education (NRF-2018R1D1A1A09081919).
\end{acknowledgements}



\bibliographystyle{spphys}       

\end{document}